\documentclass[10pt,letterpaper]{article} 
\usepackage{opex3}

\begin{document}

\title{Complex $\textbf{k}$ band diagrams of 3D metamaterial/photonic crystals.}

\author{Chris Fietz$^1$, Yaroslav Urzhumov$^2$ and Gennady Shvets$^{1*}$}
\address{$^1$Dept. of Physics, The University of Texas at Austin, Austin TX 78712,
USA \\ $^2$Center for Metamaterials and Integrated Plasmonics, Pratt School of Engineering,
Duke University, Durham, North Carolina 27708,
USA}\label{add_Duke}

\email{$^*$gena@physics.utexas.edu}

\begin{abstract}
A finite element method (FEM) for solving the complex valued $\textbf{k}(\omega)$ vs. $\omega$ dispersion curve of a 3D
metamaterial/photonic crystal system is presented.  This 3D method is a generalization of a previously reported 2D eigenvalue method~\cite{Marcelo_07,Engstrom_09}. This method is particularly convenient for analyzing periodic systems containing dispersive (e.g., plasmonic) materials, for computing isofrequency surfaces in the $\textbf{k}$-space, and for calculating the decay length of the evanescent waves. Two specific examples are considered: a photonic crystal comprised of dielectric spheres and a plasmonic fishnet structure. Hybridization and avoided crossings between Mie resonances and propagating modes are numerically demonstrated. Negative index propagation of four electromagnetic
modes distinguished by their symmetry is predicted for the plasmonic fishnets. By calculating the isofrequency contours, we also demonstrate that the fishnet structure is a hyperbolic medium.
\end{abstract}


\section{Introduction}
The numerical simulation of electromagnetic fields inside both metamaterial and photonic crystals is an important tool for
analyzing these periodic structures.  In particular, the eigenmodes of crystals, defined as freely propagating waves not
coupled to external currents, are often of the most interest. The conventional method~\cite{Joannopoulos_08,Sakoda_04} of
numerically solving for crystal eigenmodes is to define the geometry of the unit cell of the crystal of interest and the
differential equation that the fields must obey in this geometry and then impose Bloch periodic boundary conditions.  The partial differential equation problem is then discretized using one of the many standard methods (finite element, finite integral, finite difference, etc.) thereby turning it into an algebraic eigenvalue problem with a finite number of degrees of freedom and the frequency $\omega$ as the eigenvalue. This finite sized eigenvalue problem is then solved numerically. An important detail of this method is that the Bloch wavenumber $\textbf{k}$ is chosen beforehand, and the frequency is then computed as a function of the wavenumber, yielding the dispersion curves $\omega=\omega(\textbf{k})$. This is the most commonly used method for calculating dispersion curves of the electromagnetic waves propagating in photonic crystals or in closely related metamaterial crystals.

There are however, many instances where it is more convenient to specify the frequency $\omega$ and solve for the wavenumber as a function of frequency: $\textbf{k}=\textbf{k}(\omega)$.  At least four such instances can be identified. First, metamaterials often contain dispersive materials such as metals, where the dielectric function strongly depends on the frequency on the wave. In this case, the eigenfrequency problem becomes a nonlinear eigenvalue problem and must be solved iteratively~\cite{Shvets_PRL04}.  In contrast, when solving for the wavenumber as a function of frequency, the resulting eigenvalue problem only needs to be solved once.  Second, it is often useful to solve for the wavenumber as the eigenvalue because of the information contained in the complex wavenumber including the decay lengths of the electromagnetic modes (either due to finite dissipation or because of the evanescent nature of the mode) and the figure of merit~\cite{Brueck_OptExp05} of negative index modes.  Third, in the majority of experiments electromagnetic fields inside metamaterial/photonic crystals are excited by external sources producing time-harmonic fields with real valued frequencies. A complex wavenumber eigenvalue simulation provides the correct field distribution in the photonic crystal relevant to such an experiment. Fourth, this approach provides a very natural way of calculating the so-called isofrequency surfaces corresponding to $\omega(\textbf{k})=const$, where $\omega$ is real.  Isofrequency diagrams are fundamentally important for predicting wave refraction at the PC interfaces~\cite{notomi2000} and for calculating the density of states~\cite{mcphedran_pre04}.

There are several previously published methods on calculating $\textbf{k}(\omega)$ dispersion curves including variations of the plane-wave expansion method~\cite{Suzuki_95,Istrate_05} and diagonalizing the crystal transfer matrix~\cite{Ward_95,Li_03}.  A method of solving for complex wavenumber dispersion curves using the FEM has been proposed~\cite{Marcelo_07,Engstrom_09} but only for 2D crystals.  The benefits of the complex wavenumber 2D FEM
are becoming better appreciated and use of this method is becoming more common~\cite{Zhang_09,Zhang_10,Fietz_10a,Conforti_10,Fietz_10b}. It is therefore timely to generalize this method of Complex Wavenumber Eigenvalue Simulations (CWES) from two to three dimension, which is the object of this paper.  This 3D complex wavenumber eigenvalue simulation was recently used as part of a metamaterial homogenization procedure~\cite{Pors_11}.

The basic theory behind solving for complex wavenumber eigenvalues using the FEM discretization is explained in
Sec.~\ref{sec:FEM_sec}. The underlying field equations for the electric and magnetic fields and the necessary boundary conditions are discussed. In Sec.~\ref{photo_sec} we apply this method to a 3D photonic crystal consisting of non-dispersive dielectric spheres. The photonic band structure for electromagnetic waves propagating both parallel and obliquely to the principal axes is calculated, and different types of modes (transverse and longitudinal) are identified.  In Sec.~\ref{fishnet_sec} we calculate the dispersion curves for a negative index fishnet metamaterial~\cite{Valentine_08} and demonstrate the existence of four distinct negative index modes.
We also calculate the two-dimensional isofrequency contours for the least damped transverse mode and demonstrate from the shape of the isofrequency contours that this transverse mode is hyperbolic.

\section{The finite element eigenvalue problem}\label{sec:FEM_sec}

\subsection{The field equation}
In this section we present the FEM formulation for solving for the magnetic field.  Electromagnetic wave propagation is described by the Maxwell equations which can be rearranged into a wave equation for either the electric field $\textbf{E}$ or the magnetic field $\textbf{H}$.  The wave equation for the magnetic field is

\begin{equation}\label{Wave_eq}
\displaystyle\nabla \times \left(\frac{1}{\epsilon} \nabla \times \textbf{H}\right) - \mu\frac{\omega^2}{c^2} \textbf{H} = 0.
\end{equation}

\noindent Here $\epsilon(\textbf{x})$ and $\mu(\textbf{x})$ are the microscopic permittivity and permeability of the
metamaterial/photonic crystal of interest.  Due to the periodic nature of the crystal both are are assumed to be scalar functions periodic in the crystal lattice.  According to Bloch's theorem~\cite{Joannopoulos_08,Sakoda_04} the magnetic field can be represented as the product of a periodic function and
an exponential factor

\begin{equation}\label{Bloch_H}
\textbf{H}(\textbf{x}) = \textbf{u}(\textbf{x}) \exp[\mathrm{i}(\omega t - \textbf{k}\cdot\textbf{x})],
\end{equation}

\noindent where $\omega$ is the frequency of the wave and $\textbf{k}$ is the wavevector of the Bloch-Floquet wave.
$\textbf{u}(\textbf{x})$ is a vector function which is periodic in the crystal lattice.  By inserting Eq.~(\ref{Bloch_H}) into Eq.~(\ref{Wave_eq}) we obtain an equivalent field equation for $\textbf{u}$

\begin{equation}\label{Motion}
\displaystyle\frac{k^2}{\epsilon}\textbf{u}-\frac{\textbf{k}}{\epsilon}\left(\textbf{k}\cdot\textbf{u}\right) - i\textbf{k}\times\left(\frac{1}{\epsilon}\nabla\times\textbf{u}\right) - i\nabla\times\left(\frac{1}{\epsilon}\textbf{k}\times\textbf{u}\right) + \nabla\times\left(\frac{1}{\epsilon}\nabla\times\textbf{u}\right) - \mu\frac{\omega^2}{c^2}\textbf{u} = 0,
\end{equation}

\noindent which can be interpreted as an eigenvalue problem and solved for the Bloch wavenumber $\textbf{k}$ as the eigenvalue. The spatial profile of the eigenmode $\textbf{u}(\textbf{x})$ is also recovered providing the magnetic field profile according to Eq.~(\ref{Bloch_H}) and the electric field profile according to 

\begin{equation}\label{Bloch_E}
\displaystyle\textbf{E}(\textbf{x})=\frac{1}{\mathrm{i}\epsilon\omega/c}\nabla\times\textbf{H}=\frac{1}{\mathrm{i}\epsilon\omega/c}\left(-\mathrm{i}\textbf{k}\times\textbf{u}+\nabla\times\textbf{u}\right)\exp[\mathrm{i}(\omega t - \textbf{k}\cdot\textbf{x})]
\end{equation}

\subsection{The finite element model}\label{sec:FEM_sec_2}
There are several commercial FEM software programs (COMSOL Multiphysics by COMSOL, HFSS by Ansys, Vector Fields Opera by
Cobham Technical Services, etc.) that are available for modeling metamaterial/photonic crystals.  These commercial software packages provide a convenient graphical user interface for defining a crystal's geometry, meshing the computational domain, and visualizing the electromagnetic fields. This allows for models to be quickly developed and tested.  Of the many commercial FEM codes available, the authors of this paper are only aware of one (COMSOL Multiphysics) that allows the user to specify the field equation to be solved.  The simulation examples
and results presented here were obtained using COMSOL.  In what follows, only the most essential features of the FEM approach are reviewed; more detailed treatments can be found
elsewhere~\cite{Zimmerman_04,Jin_02,Silvester_96}.

The FEM is based on setting the integral of a so-called weak expression over the domain of interest to zero.  Doing so ensures the field equation is satisfied and also creates boundary conditions.  The weak expression corresponding to
Eq.~(\ref{Motion}) is

\begin{equation}\label{weak_form}
\begin{array}{rcl}
\mathrm{F_H}(\textbf{v},\textbf{u}) & = & \displaystyle\frac{k^2}{\epsilon}\textbf{v}\cdot\textbf{u}-\frac{1}{\epsilon}\left(\textbf{k}\cdot\textbf{v}\right)\left(\textbf{k}\cdot\textbf{u}\right)-\mathrm{i}\frac{1}{\epsilon}\textbf{v}\cdot\left[\textbf{k}\times\left(\nabla\times\textbf{u}\right)\right]-\mathrm{i}\left(\nabla\times\textbf{v}\right)\cdot\frac{1}{\epsilon}\left(\textbf{k}\times\textbf{u}\right)
\\ \\
 & & \displaystyle + \left(\nabla\times\textbf{v}\right)\cdot\frac{1}{\epsilon}\left(\nabla\times\textbf{u}\right)-\mu\frac{\omega^2}{c^2}\textbf{v}\cdot\textbf{u},
\end{array}
\end{equation}

\noindent where $\textbf{v}(\textbf{x})$ is a test function. When the integral of the weak expression over the unit cell $\Omega$ of the crystal is set to zero, integrating by parts gives us two separate integrals (Eq.~(\ref{int_parts})).  The first integral enforces the field equation.  The second integral is over the boundary of the domain and represents a natural boundary condition~\cite{Zimmerman_04,Jin_02},

\begin{equation}\label{int_parts}
\begin{array}{rcl}
0 & = & \displaystyle\int_{\Omega} d^3 x\ \ \mathrm{F_H}(\textbf{v},\textbf{u})
\\ \\
& = & \displaystyle\int_{\Omega} d^3 x \ \ \textbf{v}\cdot\left[-\frac{1}{\epsilon}\textbf{k}\times\left(\textbf{k}\times\textbf{u}\right)-\mathrm{i}\frac{1}{\epsilon}\textbf{k}\times\left(\nabla\times\textbf{u}\right)-\mathrm{i}\nabla\times\left(\frac{1}{\epsilon}\textbf{k}\times\textbf{u}\right)\right.
\\ \\ & & \left.+\displaystyle\nabla\times\left(\displaystyle\frac{1}{\epsilon}\nabla\times\textbf{u}\right)-\mu\frac{\omega^2}{c^2}\textbf{u} \right]
\\ \\
& & + \displaystyle\oint_{\partial\Omega} dA\ \textbf{v}\cdot \left[\hat{\textbf{n}}\times\frac{1}{\epsilon}\left(-\mathrm{i}\textbf{k}\times\textbf{u}+\nabla\times\textbf{u}\right)\right],
\end{array}
\end{equation}

\noindent where $\hat{\textbf{n}}$ is the vector normal to the boundary.  On an external boundary, the natural boundary condition enforced by the integral in Eq.~(\ref{int_parts}) over the boundary $\partial\Omega$ forces the expression
$\hat{\textbf{n}}\times\left(-\mathrm{i}\textbf{k}\times\textbf{u}+\nabla\times\textbf{u}\right)/\epsilon$ to be equal to zero.  Recalling Eq.~(\ref{Bloch_E}) we note that this simply enforces the boundary condition $\hat{\textbf{n}}\times\textbf{E}=0$.  This is known as the perfect electric conductor or PEC boundary condition. This natural boundary condition is the default if no other boundary condition is enforced.  On an internal boundary within the unit cell the surface integrals over each side of the boundary must be equal to each other.  The effect is that the tangential components of the electric field must be continuous across the internal boundary or
$\hat{\textbf{n}}\times\textbf{E}^+=\hat{\textbf{n}}\times\textbf{E}^-$ where $\textbf{E}^+$ and $\textbf{E}^-$ are the electric fields on opposite sides of the internal boundary.

The periodicity of $\textbf{u}$ is enforced by imposing periodic boundary conditions on the exterior boundaries of the unit cell. In COMSOL, these periodic boundary conditions override the natural boundary condition.  However, if a PEC boundary condition is desired inside the unit cell (e.g., on the surface of a metal inclusion) this can be accomplished by removing the subdomain representing the metal inclusion.  Now only the exterior side of the metal boundary remains and the natural boundary condition forces the tangential electric fields to zero on this boundary.

If a perfect magnetic conductor or PMC boundary condition ($\hat{\textbf{n}}\times\textbf{H}$) is desired while solving for the magnetic field, this can be enforced with constraints~\cite{Zimmerman_04} on the tangential magnetic field on the boundary.

In order to turn Eq.(~\ref{int_parts}) into an eigenvalue problem, the three degrees of freedom that comprise the Bloch wavevector $\textbf{k}$ must be reduced to one by restricting two degrees of freedom.  This is accomplished by setting
$\textbf{k}=\textbf{k}_0+\lambda\hat{\textbf{k}}_n$ where $\lambda$ will be the eigenvalue solved for, $\textbf{k}_0$ is an offset vector and $\hat{\textbf{k}}_n$ is a unit vector ($\hat{\textbf{k}}_n\cdot\hat{\textbf{k}}_n=1$) that defines the direction of the complex wavenumber eigenvalue $\lambda$.  The FEM turns the weak form and accompanying boundary conditions into an algebraic problem, in this case a quadratic eigenvalue problem~\cite{Tisseur_01}:

\begin{equation}
\mathrm{A}\vec u+\lambda\mathrm{B}\vec u+\lambda^2\mathrm{C}\vec u=0,
\end{equation}

\noindent where $\mathrm{A}$, $\mathrm{B}$ and $\mathrm{C}$ are $\mathrm{N}\times \mathrm{N}$ matrices and $\vec u$ is an
$\mathrm{N}\times 1$ solution vector.  $\mathrm{N}$ is the number of degrees of freedom of the discretized system.  Terms in the weak form (Eq.~(\ref{weak_form})) that are zero, first and second order in $\lambda$ contribute to the $\mathrm{A}$, $\mathrm{B}$ and $\mathrm{C}$ matrices respectively.  This quadratic eigenvalue problem can be recast in the form of a generalized eigenvalue equation

\begin{equation}
\left(\!\!\!\begin{array}{cc}
\mathrm{A} & \mathrm{B} \\ 0 & 1
\end{array}\!\!\!\right)
\left(\!\!\!\begin{array}{c}
\vec u \\ \lambda \vec u
\end{array}\!\!\!\right) = \lambda
\left(\!\!\!\begin{array}{cc}
0 & -\mathrm{C} \\ 1 & 0
\end{array}\!\!\!\right)
\left(\!\!\!\begin{array}{c}
\vec u \\ \lambda \vec u
\end{array}\!\!\!\right).
\end{equation}
When using COMSOL to solve the FEM problem, this linearization is performed automatically during the solution
stage.

\subsection{Electric field formulation}
The previous discussion focused on solving for the magnetic field $\textbf{H}$ or rather  the periodic function $\textbf{u}$ equal to the magnetic field with the exponential Bloch factor removed. This is especially convenient when an inclusion requires a PEC boundary condition since that is the natural boundary condition when solving for $\textbf{H}$.  However, solving for the electric field is very similar to solving for the magnetic field.  The wave equation for the electric field for a free wave is

\begin{equation}
\displaystyle\nabla \times \left(\frac{1}{\mu} \nabla \times \textbf{E}\right) - \epsilon\frac{\omega^2}{c^2} \textbf{E} = 0.
\end{equation}

\noindent As before, we replace the electric field with a periodic vector field times an exponential factor

\begin{equation}
\textbf{E}(\textbf{x}) = \textbf{u}(\textbf{x}) \exp[\mathrm{i}(\omega t - \textbf{k}\cdot\textbf{x})],
\end{equation}

\noindent producing the new field equation

\begin{equation}
\displaystyle\frac{k^2}{\mu}\textbf{u}-\frac{\textbf{k}}{\mu}\left(\textbf{k}\cdot\textbf{u}\right) - i\textbf{k}\times\left(\frac{1}{\mu}\nabla\times\textbf{u}\right) - i\nabla\times\left(\frac{1}{\mu}\textbf{k}\times\textbf{u}\right) + \nabla\times\left(\frac{1}{\mu}\nabla\times\textbf{u}\right) - \epsilon\frac{\omega^2}{c^2}\textbf{u} = 0.
\end{equation}

\noindent The corresponding weak form for this field equation is

\begin{equation}
\begin{array}{rcl}
\mathrm{F_E}(\textbf{v},\textbf{u}) & = & \displaystyle\frac{k^2}{\mu}\textbf{v}\cdot\textbf{u}-\frac{1}{\mu}\left(\textbf{k}\cdot\textbf{v}\right)\left(\textbf{k}\cdot\textbf{u}\right)-\mathrm{i}\frac{1}{\mu}\textbf{v}\cdot\left[\textbf{k}\times\left(\nabla\times\textbf{u}\right)\right]-\mathrm{i}\left(\nabla\times\textbf{v}\right)\cdot\frac{1}{\mu}\left(\textbf{k}\times\textbf{u}\right)
\\ \\
& & \displaystyle+\left(\nabla\times\textbf{v}\right)\cdot\frac{1}{\mu}\left(\nabla\times\textbf{u}\right)-\epsilon\frac{\omega^2}{c^2}\textbf{v}\cdot\textbf{u},
\end{array}
\end{equation}

\noindent which is equivalent to Eq.~(\ref{weak_form}) if $\epsilon$ and $\mu$ are interchanged.  Integrating this weak form over the crystal unit cell by parts and setting its value to zero again gives produces two integrals, one enforcing the field equation and a surface integral enforcing the boundary condition $\hat{\textbf{n}}\times\textbf{H}=0$.  Thus the PMC boundary condition is the natural boundary condition when solving for the electric field.

\section{Example: Dielectric Photonic Crystal}\label{photo_sec}
\begin{figure}[!t]
\begin{center}
\includegraphics[width=\textwidth]{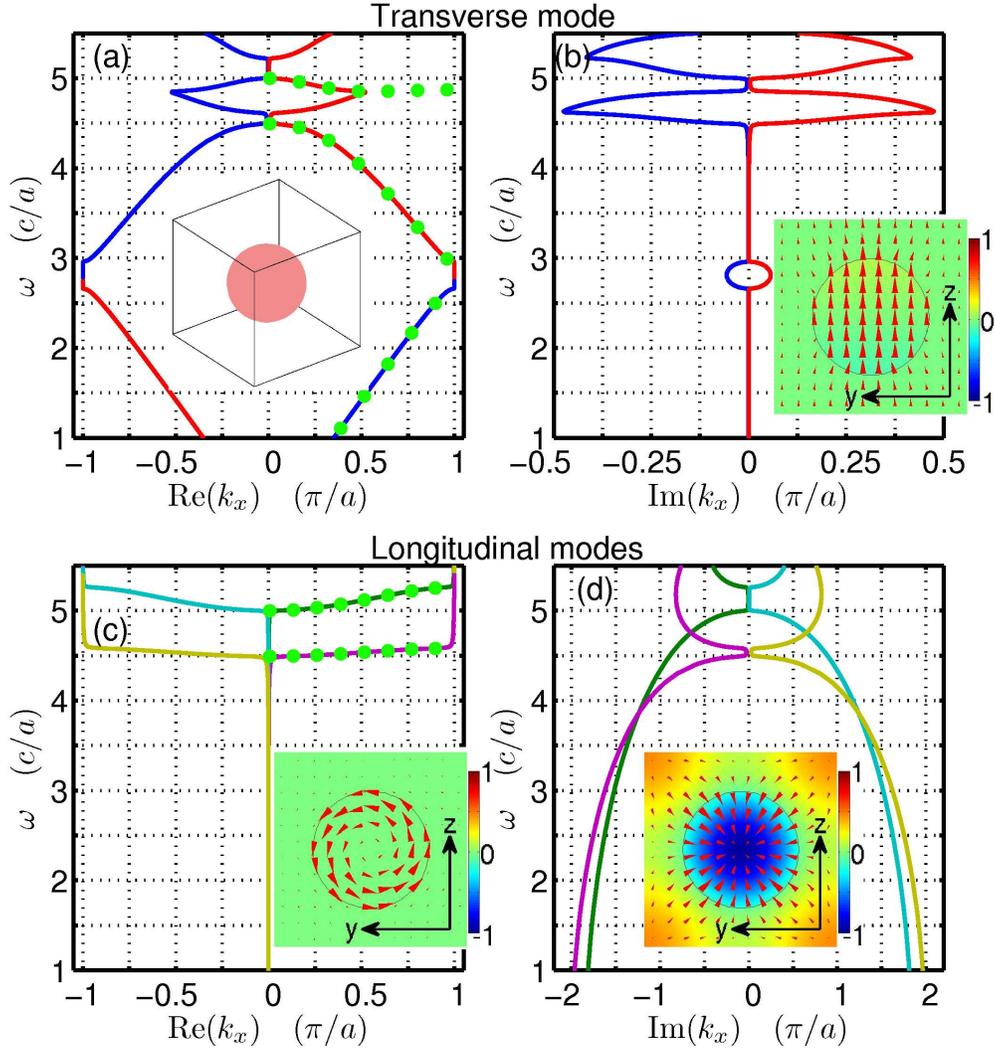}
\end{center}
\caption{Complex $\textbf{k}$ dispersion curves and field profiles for eigenmodes of the photonic crystal pictured in the inset assuming $\textbf{k}_0=0$ and $\hat{\textbf{k}}_n=\hat{\textbf{x}}$.  (a) Real part of $k_x(\omega)$ for a transversely polarized mode and a diagram of the crystal unit cell.  (b) Imaginary part of $k_x(\omega)$ for a transversely polarized mode and a field profile for the $\hat{\textbf{z}}$ polarized transverse mode. There are two transverse modes, $\hat{\textbf{y}}$ and $\hat{\textbf{z}}$ electrically polarized, which are degenerate.  (c) Real part of $k_x(\omega)$ for two longitudinally polarized modes and a field profile for the magnetic longitudinal mode.  (d) Imaginary part of $k_x(\omega)$ for two longitudinally polarized modes and a field profile for the electric longitudinal mode.  The longitudinal mode with the passband near $\omega=4.5c/a$ is magnetically polarized in the $\hat{\textbf{x}}$ direction and the longitudinal mode with
the passband near $\omega=5c/a$ is electrically polarized in the $\hat{\textbf{x}}$ direction.  The longitudinal modes correspond to Mie's dipole resonances. For all dispersion curves the dotted lines are the result of a conventional $\omega(\textbf{k})$ eigenvalue simulation.  For all field profiles the frequency is $\omega=2c/a$ with arrows representing $\mathrm{D}_y$ and $\mathrm{D}_z$ and color representing $\mathrm{D}_x$.}\label{sphere_1}
\end{figure}

For a demonstration of the CWES method of calculating complex $\textbf{k}$ dispersion curves we use a simple photonic crystal as an example.  The unit cell, pictured in Fig.~\ref{sphere_1}, is a cube with a dielectric sphere at the center surrounded by vacuum. The sphere has a radius of $0.3a$, where $a$ is the lattice constant of the cubic array, and a permittivity of $\epsilon=5-\mathrm{i}0.01$.

As mentioned in Sec.~\ref{sec:FEM_sec_2}, it is necessary to restrict two of the three degrees of freedom of the Bloch
wavevector $\textbf{k}$.  There are many possible ways to do this. As the first example, we calculate the dispersion curves corresponding to the propagation along a principal axis. To simulate this we set $\textbf{k}_0=0$ and
$\hat{\textbf{k}}_n=\hat{\textbf{x}}$.  The results of this eigenvalue simulation for the frequency range $1c/a\leq\omega\leq5.5c/a$ are plotted in Fig.~\ref{sphere_1} as $\omega$ vs. $k_x=\hat{\textbf{x}}\cdot\textbf{k}=\lambda$.

For clarity, we have only plotted the three least evanescent (i.e. possessing the smallest values of $\mathrm{Im}(k_z)$) eigenmodes. The three eigenmodes in Fig.~\ref{sphere_1} are described as either transverse or longitudinal according to their polarization. The symmetry of the dispersion curves is such that for every solution $\textbf{k}(\omega)$ there is also the solution $-\textbf{k}(\omega)$ indicating that this is a reciprocal crystal.  The dispersion curves for the transverse modes plotted in Fig.~\ref{sphere_1} in fact represent two polarization-degenerate modes because of the symmetry of the crystal. The longitudinal mode with the passband near $\omega=4.5c/a$ is magnetically polarized in the
$\hat{\textbf{x}}$ direction making it a magnetic bulk plasmon. The longitudinal mode with the passband near $\omega=5c/a$ is electrically polarized in the $\hat{\textbf{x}}$ direction and is an electric bulk plasmon.  The field profiles of both longitudinal modes indicate that the passbands correspond to Mie's resonances of the dielectric sphere~\cite{Bohren_98}.

\begin{figure}[!t]
\begin{center}
\includegraphics[width=\textwidth]{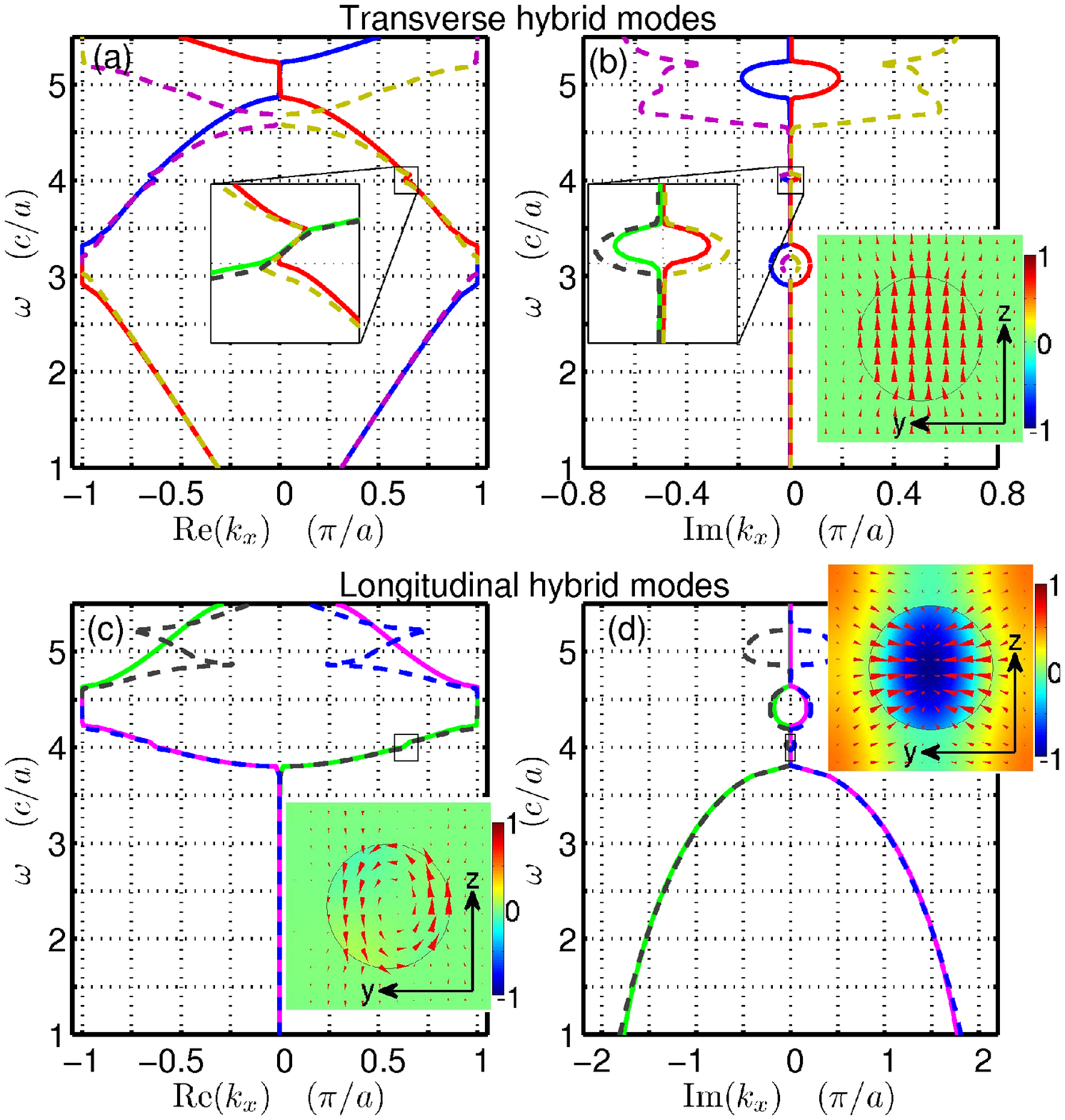}
\end{center}
\caption{Complex wavenumber dispersion curves and field profiles for eigenmodes of the photonic crystal pictured in
Fig.~\ref{sphere_1}(a) assuming $\textbf{k}_0=\omega/c\sin(\pi/6) \hat{\textbf{y}}$ and $\hat{\textbf{k}}_n=\hat{\textbf{x}}$. Modes excited by p or s polarized incident light are plotted with solid or dashed lines respectively.  (a) Real part of $k_x(\omega)$ for two transverse hybrid modes and a expanded view of the avoided crossing in $\mathrm{Re}(k_x)$ space.  (b) Imaginary part of $k_x(\omega)$ for two transverse hybrid modes, an expanded view of the avoided crossing in $\mathrm{Im}(k_x)$ space (plotting the same modes as the expanded view in $\mathrm{Re}(k_x)$ space), and a field profile for the $\mathrm{E}_z$ polarized transverse hybrid mode.  (c) Real part of
$k_x(\omega)$ for two longitudinal hybrid modes and a field profile for the magnetic longitudinal hybrid mode.  (d) Imaginary part of $k_x(\omega)$ for two longitudinal hybrid modes and a field profile for the electric longitudinal hybrid mode.  The magnetic longitudinal hybrid mode is excited by s polarized incident light and the electric longitudinal hybrid mode is excited by p polarized incident light.  For all field profiles the frequency is $\omega=2c/a$ with arrows representing $\mathrm{D}_y$ and $\mathrm{D}_z$ and color representing $\mathrm{D}_x$.}\label{sphere_2}
\end{figure}

The transverse mode dispersion curve has a band in the approximate frequency range $4.6c/a<\omega<4.8c/a$ with a large value of $\mathrm{Im}(k_x)$, indicating it is an evanescent band, but a $\mathrm{Re}(k_x)$ that is equal to neither $0$ nor $\pi/a$ as is typical of $\omega(\textbf{k})$ dispersion curves.  As described in Refs.~\cite{Marcelo_07,Tisseur_01} for a quadratic eigenvalue problem with hermitian matrices (corresponding to a lossless crystal) the eigenvalues must always be real or come in complex-conjugate pairs.  The dielectric photonic crystal under consideration has very low loss, so this lossless condition approximately holds true for the dispersion curves in Fig.~\ref{sphere_1}.  The transverse band in the $4.6c/a<\omega<4.8c/a$ frequency band is one half of a complex conjugate pair, the other half is a transverse doubly degenerate mode not shown here.  At the frequency of $\omega\approx4.8c/a$ the two modes that make up this complex conjugate pair both enter a passband and split, the plotted mode going to the $\Gamma$ point and the unplotted mode going to the band edge (this unplotted mode corresponds to the dotted lines from the $\omega(\textbf{k})$ simulation).  Note that in this passband there are two pairs of propagating doubly polarization degenerate modes or four propagating modes in total.

The transverse eigenmodes plotted in Fig.~\ref{sphere_1} can be excited by a plane wave normally incident onto the vacuum/photonic crystal interface if the interface is parallel to the $\textbf{y}$-$\textbf{z}$ plane. The longitudinally polarized modes could not be excited by a normally incident wave without the aid of a coupling device at the interface.  If the incident beam of light is not normal to the interface, if for example the incident beam has a wavenumber laying in the $\textbf{x}$-$\textbf{y}$ plane but at a $30^{\circ}$ angle from normal then a different set of eigenmodes will be excited at the interface.  To simulate these excited eigenmodes we set $\textbf{k}_0=\omega/c\sin(\pi/6)\hat{\textbf{y}}$ and $\hat{\textbf{k}}_n=\hat{\textbf{x}}$ and solve the resulting eigenvalue problem.  The resulting photonic band structure is plotted in Fig.~\ref{sphere_2}.

The eigenmodes in Fig.~\ref{sphere_2} are roughly split into predominantly-transverse (transverse hybrid) and
predominantly-longitudinal (longitudinal hybrid) modes. The hybridization between the transverse and longitudinal modes is
caused by the finite symmetry-breaking $k_y$. Both the transverse and longitudinal hybrid modes in Fig.~\ref{sphere_2} can be characterized by the polarization of the incident light that couples to them. For a p polarized incident beam (electric field in the $\textbf{x}$-$\textbf{y}$ plane) the p polarized eigenmode is excited (plotted in Fig.~\ref{sphere_2} with solid lines) and for an s polarized incident beam (electric field in the $\hat{\textbf{z}}$ direction) the s polarized eigenmode is excited (plotted in Fig.~\ref{sphere_2} with dashed lines).

At the frequency $\omega\approx4c/a$ the transverse hybrid modes and the longitudinal hybrids modes appear to cross in a propagating band.  An expanded view of this region in Fig.~\ref{sphere_2} plotting both transverse and longitudinal hybrid modes shows that the apparent crossing actually occurs in a band gap.  Viewed in complex $k_x$ space it is clear that this is actually an avoided crossing and in the band gap the transverse and longitudinal hybrid modes form complex conjugate pairs~\cite{Tisseur_01,Marcelo_07}.

We see that even for a simple photonic crystal the CWES method of calculating the dispersion curve produces rich and complex results.  In particular, it is not possible to solve for evanescent eigenmodes using the conventional $\omega(\textbf{k})$ method for calculating dispersion curves.

\section{Example: Negative Index Plasmonic
Fishnet}\label{fishnet_sec}

\begin{figure}[h]
\begin{center}
\includegraphics[height=4cm]{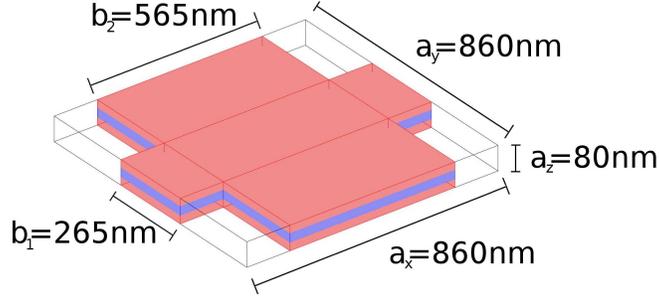}
\end{center}
\caption{The fishnet metamaterial from Ref.~\cite{Valentine_08}. The lattice constants of the unit cell are $a_x=a_y=860nm$ and $a_z=80nm$.  The fishnet is made up of alternating layers of Ag with a thickness of $30nm$ and $\mathrm{MgF}_2$ with a thickness of $50nm$.  The widths of the the crisscrossing fishnet strips are $b_1=265nm$ and $b_2=565nm$.}\label{fishnet_1}
\end{figure}

The second example highlighting the versatility of the CWES method is a plasmonic negative index metamaterial (NIM) shown in Fig.~\ref{fishnet_1}. Because this metamaterial contains dispersive (plasmonic) components, it is even more convenient to use this method because the dielectric permittivities of metals are tabulated for real-valued frequencies. Recent
experiments~\cite{Valentine_08} demonstrated that the so-called fishnet metamaterial supports a negative index eigenmode for near infrared wavelengths of about $\lambda_0\approx1.7\mu m$. The dimensions and composition of the unit cell taken from Ref.~\cite{Valentine_08} are shown in Fig.~\ref{fishnet_1}. The fishnet metamaterial is made of alternating layers of Ag and $\mathrm{MgF}_2$ with thicknesses of $30nm$ and $50nm$ respectively.  This layered structure was milled with a focused ion beam into crisscrossing strips with widths of $265nm$ and $565nm$.  The crystal lattice constants are $a_x=a_y=860nm$ and $a_z=80nm$.  For the dielectric response of the Ag we used the Drude model $\epsilon_{Ag}=1-\omega_p^2/(\omega(\omega-\mathrm{i}\gamma))$ with the same parameters cited in Ref.~\cite{Valentine_08}:
$\omega_p=9eV$ and $\gamma=0.054eV$.  It should be noted that in Ref.~\cite{Valentine_08} the scattering frequency $\gamma$ was increased by a factor of three from that of standard bulk Ag to account for additional scattering losses.  For the dielectric response of $\mathrm{MgF}_2$, no values were provided in the original paper so we used an oscillator model of the dielectric permittivity from Ref.~\cite{Siqueiros_88}.

\begin{figure}[!t]
\begin{center}
\includegraphics[width=\textwidth]{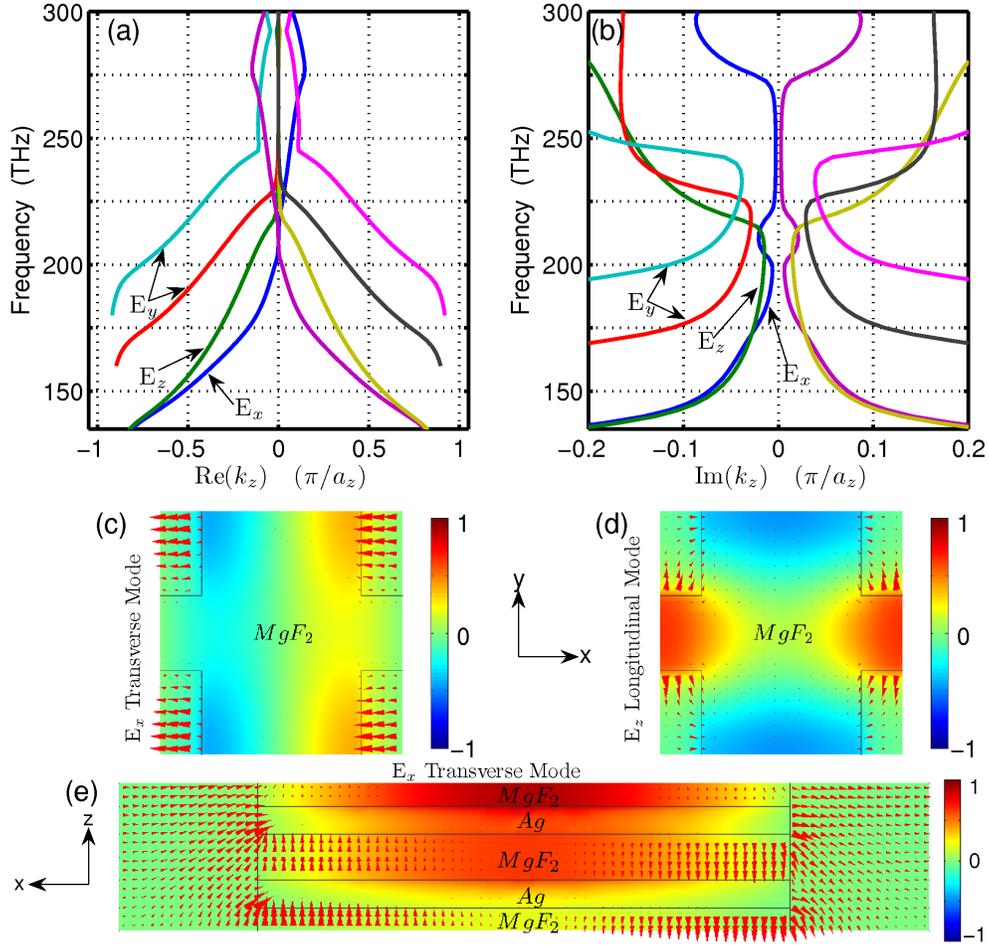}
\end{center}
\caption{TOP: Real (a) and imaginary (b) parts of $k_z(\omega)$ for several eigenmodes of the fishnet metamaterial shown in Fig.~\ref{fishnet_1}.  In addition to the transverse mode electrically polarized in the $\hat{\textbf{x}}$ direction labeled $\mathrm{E}_x$ which was identified in Ref.~\cite{Valentine_08} we see two other transverse modes electrically polarized in the $\hat{\textbf{y}}$ direction labeled $\mathrm{E}_y$ and a longitudinal mode electrically polarized in the $\hat{\textbf{z}}$ direction labeled $\mathrm{E}_z$.  MIDDLE: Field profiles for the transverse $\mathrm{E}_x$ mode (c) and the longitudinal $\mathrm{E}_z$ mode (d) on a cross-section laying on the $\textbf{x}$-$\textbf{y}$ plane in the middle of the $\mathrm{MgF}_2$ layer.  Arrows represent in-plane electric field and color represents the $\mathrm{E}_z$ field.  BOTTOM: (e) Field profile of the transverse $\mathrm{E}_x$ mode on a cross-section laying on the $\textbf{x}$-$\textbf{z}$ plane halfway between
two thin $Ag$-$MgF_2$ strips.  Arrows represent in-plane electric field and color represents the $\mathrm{H}_y$ field.  For all field profiles the frequency is $175 THz$ and each region is labeled $Ag$ or $MgF_2$ according the the material of the region.  Unlabeled regions are vacuum.}\label{fishnet_2}
\end{figure}



The negative index mode reported in Ref.~\cite{Valentine_08} propagates in the $\hat{\textbf{z}}$ direction. Therefore, we have applied CWES to the specific case of $\textbf{k}_0=0$ and $\hat{\textbf{k}}_n=\hat{\textbf{z}}$.  The resulting dispersion curves are plotted in Fig.~\ref{fishnet_2} for the $100 THz < \omega/2\pi < 300 THz$ frequency range. For clarity we have only plotted the four eigenmodes that have the smallest values of $\mathrm{Im}(k_z)$ and therefore the longest propagation lengths. The negative index mode found in Ref.~\cite{Valentine_08} is labeled $\mathrm{E}_x$ to indicate that it is transversely polarized with the electric field pointing in the $\hat{\textbf{x}}$ direction.  We can see that it is a negative index mode because (according to the convention where fields are Bloch periodic with the exponential factor $\exp[\mathrm{i}(\omega t - \textbf{k}\cdot\textbf{x})]$) a negative index mode should have a wavenumber whose real and imaginary parts have the same sign. In Fig.~\ref{fishnet_2}, that describes the $\mathrm{E}_x$ mode for
frequencies below $200 THz$.  In addition, we observe three other negative index modes, though they generally have a shorter
propagation lengths than the $\mathrm{E}_x$ mode. Of these three modes, the two labeled $\mathrm{E}_y$ are transversely polarized with the electric field pointing in the $\hat{\textbf{y}}$ direction and the mode labeled $\mathrm{E}_z$ is longitudinally polarized with the electric field pointing in the $\hat{\textbf{z}}$ direction.

\begin{figure}[t]
\begin{center}
\includegraphics[width=\textwidth]{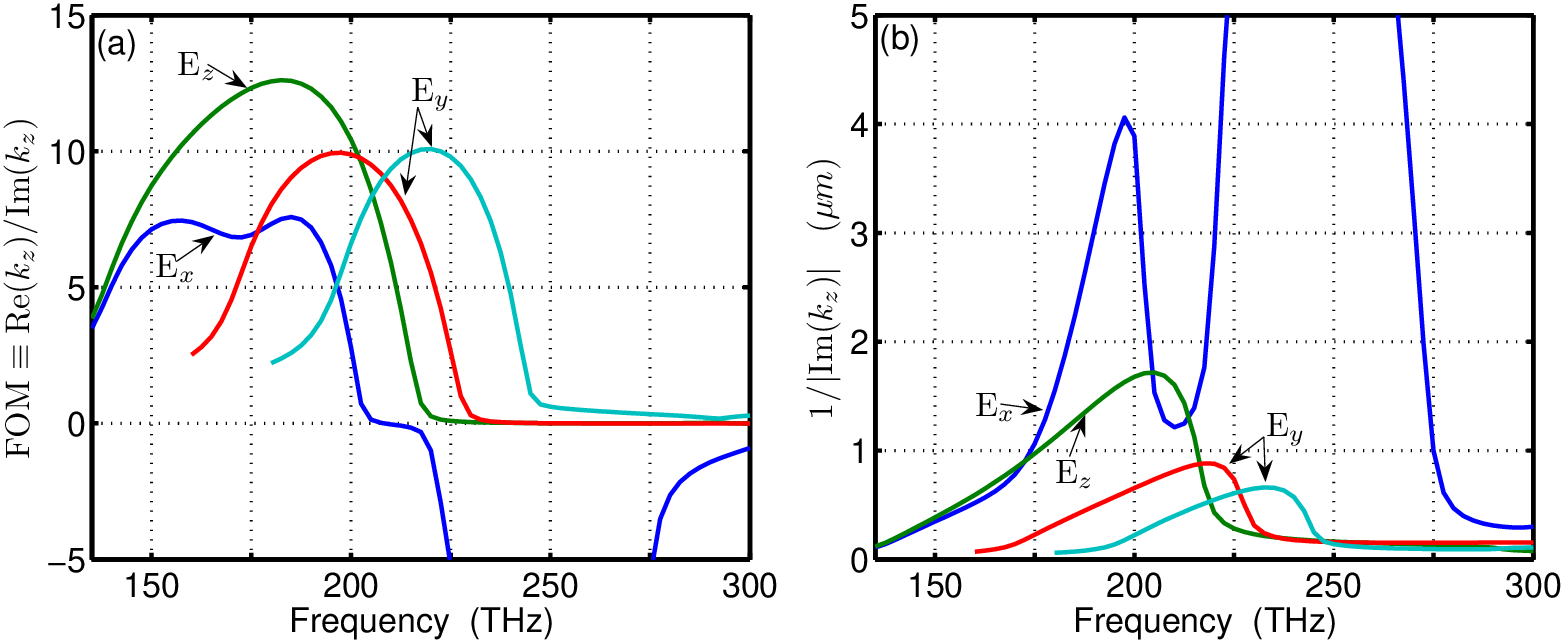}
\end{center}
\caption{The FOM (a) and propagation length (b) for the four fishnet eigenmodes identified in Fig.~\ref{fishnet_2}.  Note that though the mode labeled $E_x$ has the largest propagation length it also has the smallest FOM.  The negative value of FOM for the $E_x$ mode for frequencies above $220 THz$ indicates that it is no longer a negative index mode at these higher frequencies.}\label{fishnet_3}
\end{figure}

The figure of merit (FOM) is a number commonly used to quantify the quality of a negative index material.  It is often
defined~\cite{Brueck_OptExp05} as the ratio between the real and imaginary parts of the index of refraction.  For eigenmodes of the fishnet crystal with real $\omega$ and a complex $\textbf{k}$ pointing in the $\hat{\textbf{z}}$ direction this is equivalent to the ratio between the real and imaginary parts of $k_z$ or FOM$\equiv\mathrm{Re}(k_z)/\mathrm{Im}(k_z)$.  Without knowledge of the complex wavenumber of an eigenmode of a metamaterial, the figure of merit must be calculated indirectly.  For example, in Ref.~\cite{Valentine_08} the FOM is estimated numerically from simulated transmission through a fishnet sample. With knowledge of the complex wavenumbers of the crystal eigenmodes, it is possible to calculate the FOM directly.   The FOM for each of the four eigenmodes identified in Fig.~\ref{fishnet_2} is plotted in Fig.~\ref{fishnet_3}(a), along with the propagation lengths plotted
in Fig.~\ref{fishnet_3}(b). A positive FOM corresponds to the real and imaginary parts of $k_z$ having the same sign, therefore indicating a negative index eigenmode. We see again that all four modes are in fact negative index modes in a fairly broad $150 THz < \omega/2\pi < 200 THz$ frequency range. Somewhat unexpectedly, the mode experimentally observed in Ref.~\cite{Valentine_08} (labelled $E_x$) has the lowest FOM despite having the longest propagation length. The theoretically predicted FOM$\approx 7.5$ of the $E_x$ mode is a factor of $2$ higher than the experimentally measured FOM~\cite{Valentine_08}, which could be attributed to fabrication imperfections.

The existence of multiple negative index waves in the fishnet structure has important experimental implications. For some
frequencies (e.g., at $150 THz$) the propagation length of the longitudinal mode ($E_z$) is as long as that of the primary transverse mode ($E_x$). While the original experiments~\cite{Valentine_08} only excited the $E_x$ mode by using the light normally incident onto the $x-y$ vacuum/fishnet interface, one can envision exciting both $E_x$ and $E_z$ modes
using obliquely incident light. For example, if p-polarized light is obliquely incident with the incident wavevector laying in the $\textbf{x}$-$\textbf{z}$ plane (electric field laying in the $\textbf{x}$-$\textbf{z}$ plane), then both $E_x$ and $E_z$ modes will be launched into the fishnet with different phase velocities. Their refraction by the fishnet prism~\cite{Valentine_08} would give rise to two distinct beams. The existence of the additional strongly dispersive longitudinal modes (bulk plasmons) also suggests that the fishnet is a metamaterial with significant spatial dispersion that can have a strong effect on the surface waves at the vacuum/fishnet interface~\cite{Halevi_92,Shapiro_OptLett06}. Strong spatial dispersion is not unexpected because the fishnet's transverse period is about $\lambda/2$.

Recent theoretical~\cite{Jelinek_10} and experimental~\cite{Sorolla_OptExp08} studies demonstrated that negative index propagation in fishnet structures is not limited to the optical frequency range and can, in fact, be observed with
microwaves. The main physical distinction (excluding their appropriately scaled sizes) between microwave and optical
structures is that the metals in the former can be accurately described as PECs. On the other hand, in the optical range metals are described as being "plasmonic", which means that optical field penetration into the metal is significant. Our simulation can quantify how important the plasmonic properties of the metal are for the fishnet structure from Ref.~\cite{Valentine_08}. This is done by introducing the plasmonic parameter $T_p$~\cite{Urzhumov_08}, which is the number that characterizes the plasmonic nature of a metamaterial and it is defined as the ratio of the kinetic energy of the plasmonic electrons to the magnetic energy in the unit cell of a crystal. Strong plasmonic effects and the importance of electrostatic resonances imply $T_p \gg 1$. We find that the plasmonic parameter for the mode labelled $\mathrm{E}_x$ in Fig.~\ref{fishnet_2} at a frequency of $175 THz$ is $T_p=0.24$.  Being less than one indicates this mode is not predominately plasmonic in nature because the electromagnetic fields do not significantly penetrate into the Ag.  This is consistent with a recent demonstration of a negative index mode in a PEC fishnet structure~\cite{Sorolla_OptExp08,Jelinek_10}.

Finally, we describe another application of the CWES method: calculation of isofrequency $\omega(\textbf{k})=const$
contours in metamaterial crystals. From the early days of metamaterials reseach, isofrequency diagrams have been used as a simple visual tool for studying refraction at the vacuum/metamaterial interfaces, especially in the context of negative index propagation and negative refraction~\cite{Joanop_PRB02}. Traditionally, isofrequency diagrams are drawn using a conventional $\omega(\textbf{k})$ eigenvalue simulation~\cite{Joannopoulos_08}.  This provides no information about propagation lengths which are important in lossy metamaterials such as the fishnet. The conventional approach is also highly laborious for plasmonic fishnets because of the strong frequency dependence of the dielectric permittivities of metals. Other semi-analytic techniques used for analyzing wave propagation in highly symmetric PEC-based fishnets~\cite{Jelinek_10} do not apply to plasmonic fishnets of interest.

\begin{figure}[t]
\begin{center}
\includegraphics[width=\textwidth]{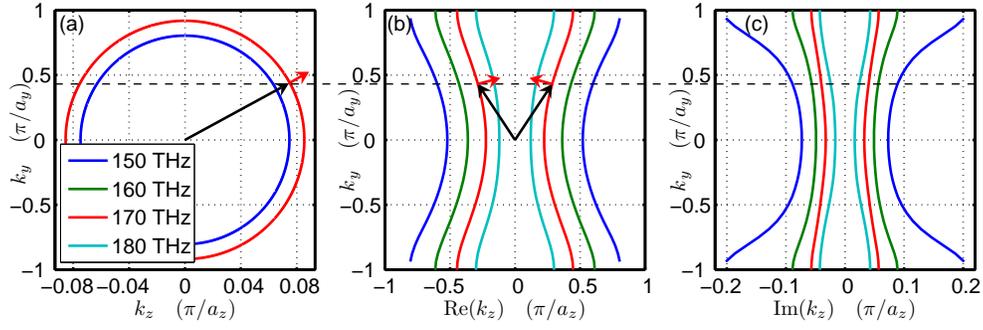}
\end{center}
\caption{(a) Isofrequency contours for vacuum plotted with respect to $k_z$ and $k_y$  (b) Isofrequency contours for the
$\mathrm{E}_x$ polarized eigenmode of the fishnet crystal plotted with respect to $\mathrm{Re}(k_z)$ and $k_y$.  (c) Isofrequency contours for the $\mathrm{E}_x$ polarized eigenmode of the fishnet crystal plotted with respect to $\mathrm{Im}(k_z)$ and $k_y$.  The black arrows indicate the direction of the phase velocity and the red arrows indicate the direction of the group velocity at a frequency of $170 THz$ and $k_y=\omega/c\sin(\pi/6)$.}\label{fishnet_4}
\end{figure}

Figure~\ref{fishnet_4} shows an isofrequency diagram calculated using the CWES approach. A single isofrequency contour was obtain by fixing the real frequency $\omega$, setting $\textbf{k}_0=k_y\hat{\textbf{y}}$ and
$\hat{\textbf{k}}_n=\hat{\textbf{z}}$, and then scanning $k_y$ from $-\pi/a_y$ to $\pi/a_y$. The eigenvalue in this simulation is a complex valued $k_z$. This procedure was repeated for several values of $\omega$ from the $150 THz$ to $180 THz$. The resulting eigenmodes can be be excited by a plane wave incident on an interface between vacuum and the fishnet crystal parallel with the $\hat{\textbf{x}}$-$\hat{\textbf{y}}$ plane if the wavevector of the incident wave is confined to the $\hat{\textbf{y}}$-$\hat{\textbf{z}}$ plane.  Because the $\hat{\textbf{y}}$ component of the incident wavevector is real valued, the $\hat{\textbf{y}}$ component of the wave excited in the fishnet crystal must also be real-valued.  $k_y = \sin{\theta} \omega/c$ where $\theta$ is the incidence angle with respect to the normal $\textbf{z}$. The eigenmode excited in the fishnet crystal decays in the $\hat{\textbf{z}}$ direction as indicated by
the imaginary part of $k_z$ plotted in Fig.~\ref{fishnet_4}.

The isofrequency contours in Fig.~\ref{fishnet_4} are hyperbolic in appearance.  We can study refraction at the interface between vacuum and the fishnet crystal by comparing the isofrequency contours of the fishnet to those of vacuum, which are also shown in Fig.~\ref{fishnet_4}.  The vacuum isofrequency contours are circular.  As can be seen in Fig.~\ref{fishnet_4} for frequency of $170 THz$ and with an incident angle of $30^{\circ}$ the component of the incident wavevector tangential to the interface ($k_y$) must be matched to the eigenmodes of the fishnet.  For $k_y=\omega/c\sin(\pi/6)$ and a frequency of $170 THz$ the isofrequency diagram shows two eigenmodes with the correct value of $k_y$ and equal and opposite values of $k_z$.  We find the correct mode by calculating the group velocity $v_g\equiv\partial\omega/\partial\mathrm{Re}(\textbf{k})$ which is by definition normal to the isofrequency contours.  In the absence of anomalous dispersion the group velocity indicates the direction of energy flow in the crystal~\cite{Brillouin_60}.  Choosing the correct solution requires us to choose the solution that has energy flowing in the positive $\hat{\textbf{z}}$ direction (i.e. away from the interface). This selects the solution with a negative $\mathrm{Re}(k_z)$. This is a negative index mode in the sense that in the $\hat{\textbf{z}}$ direction the phase velocity
and group velocity have opposite signs.

However, because the shape of the isofrequency contours is hyperbolic, the phase and group velocities in the
$\hat{\textbf{y}}$ direction have the same sign. Therefore, positive refraction at the $\hat{\textbf{x}}$-$\hat{\textbf{y}}$ vacuum/fishnet interface is expected according to Fig.~\ref{fishnet_4} despite the fact that a negative index eigenmode is excited. This does not contradict the experiment by Valentine et al. in Ref.~\cite{Valentine_08}, where the fishnet structure was cut in the shape of a prism. In that experiment, the
wave propagation through the fishnet was entirely in the $\hat{\textbf{z}}$ direction enabling the group and phase
velocities to be antiparallel. Negative refraction indeed occurs at a vacuum-fishnet interface tilted with respect to the principal axis of the crystal.

\section{Conclusion}
In conclusion, we have presented a three-dimensional realization of the Complex Wavenumber Eigenvalue Simulation (CWES) approach to calculating photonic band structure of metamaterial/photonic crystals. A detailed implementation of CWES using FEM discretization is described.  The CWES approach was applied to two periodic photonic structures: (a) photonic crystal comprised of dielectric spheres, and (b) plasmonic fishnet metamaterial supporting negative refractive index waves.  For case (a) we have used the results of CWES to identify both transverse and longitudinal modes and investigated their coupling that gives rise to avoided crossings and bandgap formation. For case (b) we have identified, for the first time, four negative-index modes of the fishnet structure and computed hyperbolic isofrequency surfaces for the least-damped transverse mode.

\section*{Acknowledgments}
We would like to acknowledge the developers of MUMPS~\cite{MUMPS}, ARPACK~\cite{ARPACK,PARPACK}, PETSC~\cite{petsc-user-ref} and SLEPC~\cite{slepc-users-manual}. Support of the staff of the Texas Advanced Computing Center is gratefully acknowledged. This work was supported by the Air Force Research Laboratory and the Office of Naval Research.


\begin{thebibliography}{10}
\newcommand{\enquote}[1]{``#1''}

\bibitem{Marcelo_07}
M.~Davanco, Y.~Urzhumov, and G.~Shvets, \enquote{The complex bloch bands of a
  2d plasmonic crystal displaying isotropic negative refraction,} Opt. Express
  \textbf{15}, 9681--9691 (2007).

\bibitem{Engstrom_09}
C.~Engstr{\"o}m, C.~Hafner, and K.~Schmidt, \enquote{Computations of lossy
  bloch waves in two-dimensional photonic crystals,} Journal of Theoretical
  Nanoscience \textbf{6}, 1--9 (2009).

\bibitem{Joannopoulos_08}
J.~D. Joannopoulos, R.~D. Meade, and J.~N. Winn, \emph{Photonic Crystals:
  Molding the Flow of Light} (Princeton University Press, New Jersey, 2008),
  2nd ed.

\bibitem{Sakoda_04}
K.~Sakoda, \emph{Optical properties of photonic crystals} (Springer, New York,
  2004), 2nd ed.

\bibitem{Shvets_PRL04}
G.~Shvets and Y.~A. Urzhumov, \enquote{Engineering the electromagnetic
  properties of periodic nanostructures using electrostatic resonances,} Phys.
  Rev. Lett. \textbf{93}, 243902 (2004).

\bibitem{Brueck_OptExp05}
S.~Zhang, W.~Fan, K.~J. Malloy, S.~R.~J. Brueck, N.~C. Panoiu, and R.~M.
  Osgood, \enquote{Near-infrared double negative metamaterials,} Optics Express
  \textbf{613}, 4922 (2005).

\bibitem{notomi2000}
M.~Notomi, \enquote{Theory of light propagation in strongly modulated photonic
  crystals: Refractionlike behavior in the vicinity of the photonic band gap,}
  Phys.~Rev.~B \textbf{62}, 10696 (2000).

\bibitem{mcphedran_pre04}
R.~C. McPhedran, L.~C. Botten, J.~McOrist, A.~A. Asatryan, C.~M. de~Sterke, and
  N.~A. Nicorovici, \enquote{Density of states functions for photonic
  crystals,} Phys.~Rev. ~E \textbf{69}, 016609 (2004).

\bibitem{Suzuki_95}
T.~Suzuki and P.~L. Yu, \enquote{Tunneling in photonic band strucures,} J. Opt.
  Soc. Am. B \textbf{12}, 804 (1995).

\bibitem{Istrate_05}
E.~Istrate, A.~A. Green, and E.~H. Sargent, \enquote{Behavior of light at
  photonic crystal interfaces,} Phys. Rev. B \textbf{71}, 195122 (2005).

\bibitem{Ward_95}
A.~J. Ward, J.~B. Pendry, and W.~J. Stewart, \enquote{Photonic dispersion
  surfaces,} J. Phys.: Condens. Matter \textbf{7}, 2217--2224 (1995).

\bibitem{Li_03}
Z.~Y. Li and L.~L. Lin, \enquote{Photonic band structure solved by a
  plane-wave--based transfer-matrix method,} Phys. Rev. E \textbf{67}, 046607
  (2003).

\bibitem{Zhang_09}
X.~Zhang, M.~Davanco, Y.~Urzhumov, and G.~Shvets, \enquote{A subwavelength
  near-infrared negative index material,} Appl. Phys. Lett. \textbf{94}, 131107
  (2009).

\bibitem{Zhang_10}
X.~Zhang, M.~Davanco, K.~Miller, T.~W. J.~C. Wu, C.~Fietz, D.~Korobkin, X.~Li,
  G.~Shvets, and S.~R. Forrest, \enquote{Interferometric characterization of a
  subwavelength near-infrared negative index metamaterial,} Opt. Express
  \textbf{18}, 17788--17795 (2010).

\bibitem{Fietz_10a}
C.~Fietz and G.~Shvets, \enquote{Current-driven metamaterial homogenization,}
  Physica B \textbf{405}, 2930--2934 (2010).

\bibitem{Conforti_10}
M.~Conforti and M.~Guasoni, \enquote{Dispersive properties of linear chains of
  lossy metal nanoparticles,} J. Opt. Soc. Am. B \textbf{27}, 1576--1582
  (2010).

\bibitem{Fietz_10b}
C.~Fietz and G.~Shvets, \enquote{Homogenization theory for simple metamaterials
  modeled as one-dimensional arrays of thin polarizable sheets,} Phys. Rev. B
  \textbf{82}, 205128 (2010).

\bibitem{Pors_11}
A.~Pors, I.~Tsukerman, and S.~I. Bozhevolnyi, \enquote{Effective constitutive
  parameters of plasmonic metamaterials: A rigorous approach,}
  arXiv:1104.2972v1  (2011).

\bibitem{Valentine_08}
J.~Valentine, S.~Zhang, T.~Zentgraf, E.~Ulin-Avila, D.~A. Genov, and G.~Bartal,
  \enquote{Three-dimensional optical metamaterial with a negative refractive
  index,} Science \textbf{455}, 376--380 (2008).

\bibitem{Zimmerman_04}
W.~B.~J. Zimmerman, \emph{Process Modelling and Simulation with Finite Element
  Methods} (World Scientific, Singapore, 2004).

\bibitem{Jin_02}
J.~Jin, \emph{The Finite Element Method in Electromagnetics} (John Wiley \&
  Sons, Inc., New York, 2002), 2nd ed.

\bibitem{Silvester_96}
P.~P. Silvester and R.~L. Ferrari, \emph{Finite elements for electrical
  engineers} (Cambridge University Press, Cambridge, 1996), 3rd ed.

\bibitem{Tisseur_01}
F.~Tisseur and K.~Meerbergen, \enquote{The quadratic eigenvalue problem,} SIAM
  Review \textbf{43}, 235--286 (2001).

\bibitem{Bohren_98}
C.~F. Bohren and D.~R. Huffman, \emph{Absorption and Scattering of Light by
  Small Particles} (John Wiley \& Sons, New York, 1998).

\bibitem{Siqueiros_88}
J.~M. Siqueiros, R.~Machorro, and L.~E. Regalado, \enquote{Determination of the
  optical constants of {M}g{F}2 and {Z}n{S} from spectrophotometric
  measurements and the classical oscillator method,} Appl. Opt. \textbf{27},
  2549--2553 (1988).

\bibitem{Halevi_92}
P.~Halevi, \emph{Spatial Dispersion in Solids and Plasmas} (Elsevier Science
  Publishers, Amsterdam, 1992).

\bibitem{Shapiro_OptLett06}
M.~A. Shapiro, G.~Shvets, J.~R. Sirigiri, and R.~J. Temkin, \enquote{Spatial
  dispersion in metamaterials with negative dielectric permittivity and its
  effect on surface waves,} Optics Letters \textbf{31}, 2051 (2006).

\bibitem{Jelinek_10}
L.~Jelinek, R.~Marques, and J.~Machac, \enquote{Fishnet metamaterials - rules
  for refraction and limits of homogenization,} Opt. Express \textbf{18},
  17940--17949 (2010).

\bibitem{Sorolla_OptExp08}
M.~Navarro-Cía, M.~Beruete, M.~Sorolla, and I.~Campillo, \enquote{Negative
  refraction in a prism made of stacked sub-wavelength hole arrays,} Opt.
  Express \textbf{16}, 560--566 (2008).

\bibitem{Urzhumov_08}
Y.~A. Urzhumov and G.~Shvets, \enquote{Optical magnetism and negative
  refraction in plasmonic metamaterials,} Solid State Communications
  \textbf{146}, 208--220 (2008).

\bibitem{Joanop_PRB02}
C.~Luo, S.~G. Johnson, J.~D. Joannopoulos, and J.~B. Pendry, \enquote{All-angle
  negative refraction without negative effective index,} Phys. Rev. B
  \textbf{65}, 201104(R) (2002).

\bibitem{Brillouin_60}
L.~Brillouin, \emph{Wave Propagation and Group Velocity} (Academic Press, New
  York, 1960).

\bibitem{MUMPS}
\emph{{MU}ltifrontal {M}assively {P}arallel {S}olver ({MUMPS} 4.9.2) {U}ser's
  guide} (Lyon, 2009).

\bibitem{ARPACK}
R.~B. Lehoucq, D.~C. Sorensen, and C.~Yang, \emph{ARPACK Users' Guide, Solution
  of Large-Scale Eigenvalue Problems by Implicitly Restarted Arnoldi Methods}
  (SIAM, Philadelphia).

\bibitem{PARPACK}
K.~J. Maschhoff and D.~C. Soerensen, \enquote{{PARPACK}: An efficient portable
  large scale eigenvalue package for distributed memory parallel
  architectures,} Lect. Notes Comp. Sci. \textbf{1184}, 478--486 (1996).

\bibitem{petsc-user-ref}
S.~Balay, J.~Brown, , K.~Buschelman, V.~Eijkhout, W.~D. Gropp, D.~Kaushik,
  M.~G. Knepley, L.~C. McInnes, B.~F. Smith, and H.~Zhang, \enquote{{PETS}c
  users manual,} Tech. Rep. ANL-95/11 - Revision 3.1, Argonne National
  Laboratory (2010).

\bibitem{slepc-users-manual}
J.~E. Roman, E.~Romero, and A.~Tomas, \enquote{{SLEPc} users manual,} Tech.
  Rep. DSIC-II/24/02 - Revision 3.1, D. Sistemas Inform\'aticos y
  Computaci\'on, Universidad Polit\'ecnica de Valencia (2010).

\end{thebibliography}
\end{document}